\documentclass[12pt]{article}

\usepackage{amssymb}
\usepackage{amsmath}

\usepackage{graphicx}

\begin{document} %


\title{Probe of axion-like particles in vector boson scattering at a muon collider}

\author{
S.C. \.{I}nan\thanks{Electronic address: sceminan@cumhuriyet.edu.tr}
\\
{\small Department of Physics, Sivas Cumhuriyet University, 58140,
Sivas, Turkey}
\\
{\small and}
\\
A.V. Kisselev\thanks{Electronic address:
alexandre.kisselev@ihep.ru} \\
{\small A.A. Logunov Institute for High Energy Physics, NRC
``Kurchatov Institute'',}
\\
{\small 142281, Protvino, Russian Federation}
}

\date{}

\maketitle

\begin{abstract}
We have examined the sensitivity to the axion-like particle (ALP)
couplings to electroweak gauge bosons in the diphoton production at
a future muon collider. The collisions at the $\mu^+ \mu^-$ energies
of 3 TeV, 14 TeV, and 100 TeV are addressed. The differential cross
sections versus the invariant mass of the final photons and total cross
section versus minimal diphoton invariant mass are presented. We
have derived the exclusion regions for the ALP-gauge boson coupling.
The obtained bounds are much stronger than the current experimental
bounds in the ALP mass region 10 GeV -- 10 TeV. The partial-wave
unitarity constraints on the ALP-gauge boson coupling are estimated.
We have shown that the unitarity is not violated in the region of
the ALP coupling studied in the present paper.
\end{abstract}

\maketitle


\section{Introduction} %

The strong CP problem  of the Standard Model (SM) can be solved by
introducing a spontaneously broken Peccei-Quinn symmetry
\cite{Peccei:1977_1,Peccei:1977_2}. As a result, a light
pseudo-Nambu-Goldstone boson, QCD axion, arises
\cite{Weiberg:1978,Wilzcek:1978}. The QCD axion is a well-motivated
candidate for the DM \cite{Preskill:1983}-\cite{Marsh:2016} which
can be produced via the vacuum misalignment mechanism
\cite{Preskill:1983,Abbott:1983} or as the decay of topological
defects \cite{Sikivie:2008}.

The axion-like particle (ALPs) are particles having interactions
similar to the axion. The origin of the ALP is expected to be
similar but without the relationship between its coupling constant
and mass. It means that the ALP mass can be treated independently of
its couplings to the SM fields. Since the ALPs are not directly
relevant for the QCD axion, heavy ALPs can be detected at colliders.
The production of the ALPs was studied in the $pp$
\cite{Beldenegro:2018}-\cite{Bonilla:2022} and heavy-ion
\cite{Beldenegro:2019}, \cite{Bauer:2017}-\cite{Knapen:2017_2}
collisions at the LHC, as well as at future colliders
\cite{Bauer:2019}-\cite{Yue:2022}, including electron-ion scattering
\cite{Liu:2021}-\cite{Davoudiasl:2021}. For a review on the axions
and ALPs, see \cite{Marsh:2016},
\cite{Kim:1987}-\cite{Irastorza:2022} and references therein.

Many ALP searches assume their strong couplings to the
electromagnetic term $F_{\mu\nu} \tilde{F}^{\mu\nu}$. One of the
most preferred processes to probe the ALP-photon coupling is a
light-by-light (LBL) scattering. The first evidence of the
subprocess $\gamma\gamma \rightarrow \gamma\gamma$ was observed by
the  ATLAS and CMS collaborations in high-energy ultra-peripheral
PbPb collisions \cite{ATLAS_ions_1}-\cite{CMS_ions}. The
phenomenology of the LBL scattering at the LHC was examined in
\cite{Enterria:2013}-\cite{Inan:2019}. In a number of papers
\cite{I_K:2020,I_K:2021_1}, \cite{I_K:2021_2,Ellis:2022} a
phenomenology of the LBL collisions at future $e^+e^-$ colliders
were presented. The search for ALPs in the $\gamma\gamma \rightarrow
a \rightarrow \gamma\gamma$ collision with proton tagging at the LHC
was given in \cite{Beldenegro:2018}.

It is a muon collider that could provide the simplest, but the most
striking signature of the existence of the ALPs
\cite{Haghighat:2021}-\cite{Han:2022}. Muon colliders were proposed
by F.~Tikhonin and G.~Budker in the late 1960's \cite{Tikhonin:1968,
Budker:1969}. Then they were actively discussed in the early 1980's
\cite{Skrinsky:1981,Neuffer:1983}. Muon colliders have a great
potential for high-energy physics since they can offer collisions
of elementary particles at very high energies. The point is that
muons can be accelerated in a ring without limitation from
synchrotron radiation compared to linear or circular
electron-positron colliders \cite{Blondel:1999}-\cite{Long:2021}.
Note, however, that getting high luminosity needs to solve a
technical problem related to the short muon lifetime at rest and the
difficulty of producing large numbers of muons in bunches with small
emittance \cite{Ankenbrandt:1999}-\cite{Boscolo:2019}.

The muon collider could provide a determination of the electroweak
couplings of the Higgs boson which is significantly better than what
is considered attainable at other future colliders
\cite{Barger:1997_2}-\cite{Costantini:2021}. Interest in designing
and building a muon collider is also based on its capability of
probing the physics beyond the SM. In a number of recent papers
searches for SUSY particles \cite{Capdevilla:2021_1}, WIMPs
\cite{Han:2021}, vector boson fusion \cite{Costantini:2020},
leptoquarks \cite{Asadi:2021}, lepton flavor violation
\cite{Bossi:2020}, and physics of $(g-2)_\mu$
\cite{Capdevilla:2021_2} at the muon colliders are presented.

In the present paper, we study the high energy production of the ALP
in the $\mu^+\mu^- \rightarrow \mu^+ \gamma\gamma \mu^-$ process
which goes via vector boson fusion subprocess $V_1V_2 \rightarrow a
\rightarrow \gamma\gamma$, where $V_{1,2}$ is $\gamma$ or $Z$, and
$a$ is a heavy ALP. The main goal is to obtain constraints on the
ALP-vector boson coupling as a function of the ALP mass at TeV and
multi-TeV muon colliders.

\section{ALP in gauge boson scattering} %

The interaction of the ALP $a$ with SM gauge bosons is described by
the Lagrangian
\begin{equation}\label{ALP_boson_lagrangian}
\mathcal{L}_\mathrm{int} = \frac{1}{2}\,\partial_\mu a
\,\partial^{\,\mu} \!a - \frac{1}{2} m_a^2 a^2 + g^2 C_{BB}
\frac{a}{\Lambda} B_{\mu\nu} \tilde{B}^{\mu\nu} + g'^2 C_{WW}
\frac{a}{\Lambda} W_{\mu\nu}^c \tilde{W}^{c,\mu\nu} \;,
\end{equation}
where $B_{\mu\nu}$ and $W_{\mu\nu}^c$ are the field strength of
$U(1)_Y$ and $SU(2)_L$, respectively, while $\tilde{B}_{\mu\nu}$ and
$\tilde{W}_{\mu\nu}^c$ are dual field strength tensors. As was
already mentioned above, the ALP mass $m_a$ and coupling $f_a$ can
be regarded as independent parameters. After electroweak symmetry
breaking, the ALP couples to the photon and $Z$ boson as
\begin{align}\label{ALP_final_lagrangian}
\mathcal{L}_a &= \frac{1}{2}\,\partial_\mu a \,\partial^{\,\mu} \!a
- \frac{1}{2} m_a^2 a^2 + g_{a\gamma\gamma} a F_{\mu\nu}
\tilde{F}^{\mu\nu} + g_{a\gamma Z}
a F_{\mu\nu} \tilde{Z}^{\mu\nu} \nonumber \\
&+ g_{aZZ} a Z_{\mu\nu} \tilde{Z}^{\mu\nu} \;,
\end{align}
Here $\tilde{F}_{\mu\nu} =
(1/2)\,\varepsilon_{\mu\nu\alpha\beta}F_{\alpha\beta}$ and
$\tilde{Z}_{\mu\nu} =
(1/2)\,\varepsilon_{\mu\nu\alpha\beta}Z_{\alpha\beta}$ are the dual
tensors, and
\begin{align}\label{ALP_boson_coupling}
g_{a\gamma\gamma} &= \frac{e^2}{\Lambda} [C_{WW} + C_{BB}]  \;,
\nonumber \\
g_{a\gamma Z} &= \frac{2e^2}{\Lambda s_w c_w} [c_w^2 C_{WW} - s_w^2
C_{BB}] \;,
\nonumber  \\
g_{aZZ} &= \frac{e^2}{\Lambda s_w^2 c_w^2} [c_w^4 C_{WW} + s_w^4
C_{BB}] \;,
\end{align}
where $s_w$ and $c_w$ are sine and cosine of the Weinberg angle,
respectively.

In what follows, we assume that the ALP couples to hypercharge
$U(1)_Y$, not to $SU(2)_L$, that corresponds to $C_{WW} = 0$. Let us
define $e^2 C_{BB}/\Lambda = 1/f_a$, then a set of the ALP couplings
takes the form
\begin{equation}\label{ALP_boson_coupling_final}
g_{a\gamma\gamma} =  \frac{1}{f_a} \;, \quad g_{a\gamma Z} =
-\frac{2s_w}{c_w}\frac{1}{f_a} \;, \quad g_{aZZ} =
\frac{s_w^2}{c_w^2}\frac{1}{f_a} \;.
\end{equation}

We also assume that the ALP has a nonzero total width
\begin{equation}\label{axion_width}
\Gamma_a =
\frac{\Gamma(a\rightarrow\gamma\gamma)}{\mathrm{Br}(a\rightarrow\gamma\gamma)}
\;,
\end{equation}
where
\begin{equation}\label{photon_axion_width}
\Gamma(a\rightarrow\gamma\gamma) = \frac{m_a^3}{4\pi f_a^2}
\end{equation}
is the ALP decay width into two photons. In general, the ALP can
also couple to fermions as $\partial^\mu a\bar{\psi}\gamma_\mu
\gamma_5\psi$ . But for $m_a \gg m_\psi$ the full width of the ALP
decay should be mainly defined by its decay to two photons. In our
upcoming calculations the ALP branching
$\mathrm{Br}(a\rightarrow\gamma\gamma)$ is considered as a free
parameter that is equal to (or less than) 1.

\begin{figure}[htb]
\begin{center}
\includegraphics[scale=0.43]{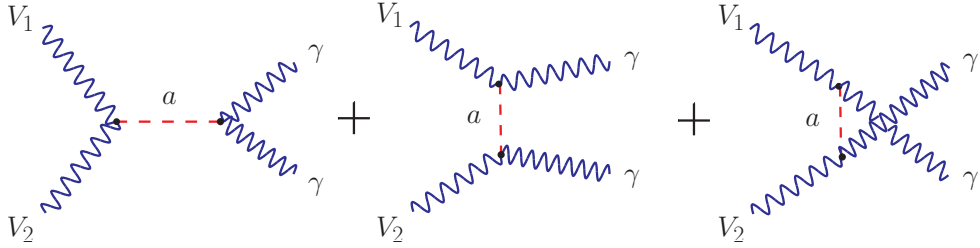}
\caption{The Feynman diagrams describing virtual production of the
axion-like particle $a$ in the collision of two vector bosons $V_1,
V_2 = \gamma$ or $Z$, with two outgoing photons.}
\label{fig:VV-gamma-gamma}
\end{center}
\end{figure}

The differential cross section of the subprocess
$V_1V_2\rightarrow\gamma\gamma$, where $V_{1,2} = \gamma$ or $Z$, is
a sum of helicity amplitudes squared
\begin{equation}\label{subprocess_cs}
\frac{d\hat{\sigma}}{d\Omega} = \frac{1}{64\pi^2 \hat{s}}
\sum_{\lambda_1,\lambda_2,\lambda_3,\lambda_4}
\!\!|M_{\lambda_1\lambda_2\lambda_3\lambda_4}^{V_1V_2}|^2 ,
\end{equation}
where $\sqrt{\hat{s}}$ is a collision energy of this subprocess, and
$\lambda_i$ are boson helicities. In its turn, each of the helicity
amplitudes in \eqref{subprocess_cs} is a sum of the ALP and SM
(electroweak) terms,
\begin{equation}\label{ALP+SM}
M = M_a + M_{\mathrm{ew}} \;.
\end{equation}
The Feynman diagrams describing $M_a$ are shown in
Fig.~\ref{fig:VV-gamma-gamma}. The explicit expressions of the ALP
helicity amplitudes of the $\gamma\gamma\rightarrow\gamma\gamma$
process can be found in \cite{Beldenegro:2018} (see also
\cite{I_K:2020}). The results of our calculations of the ALP
helicity amplitudes $M_a$ of the processes
$Z\gamma\rightarrow\gamma\gamma$ and $ZZ\rightarrow\gamma\gamma$ are
presented in Appendix~A.
%
Each of the SM amplitudes is a sum of the fermion and $W$ boson
one-loop amplitudes
\begin{equation}\label{f+W_ew}
M_{\mathrm{ew}} = M_{\mathrm{ew}}^f + M_{\mathrm{ew}}^W \;.
\end{equation}
The SM helicity amplitudes $M_{\mathrm{ew}}^f$ and
$M_{\mathrm{ew}}^W$ have been calculated for the processes
$\gamma\gamma\rightarrow\gamma\gamma$
\cite{Jikia:1994}-\cite{Gounaris:1999_2} (see also
\cite{Atag:2010}), $\gamma\gamma\rightarrow\gamma Z$
\cite{Gounaris:1999_3}, and $\gamma\gamma\rightarrow ZZ$
\cite{Gounaris:2000}.

\subsection{Differential cross section of diphoton production} 

\begin{figure}[htb]
\begin{center}
\includegraphics[scale=0.4]{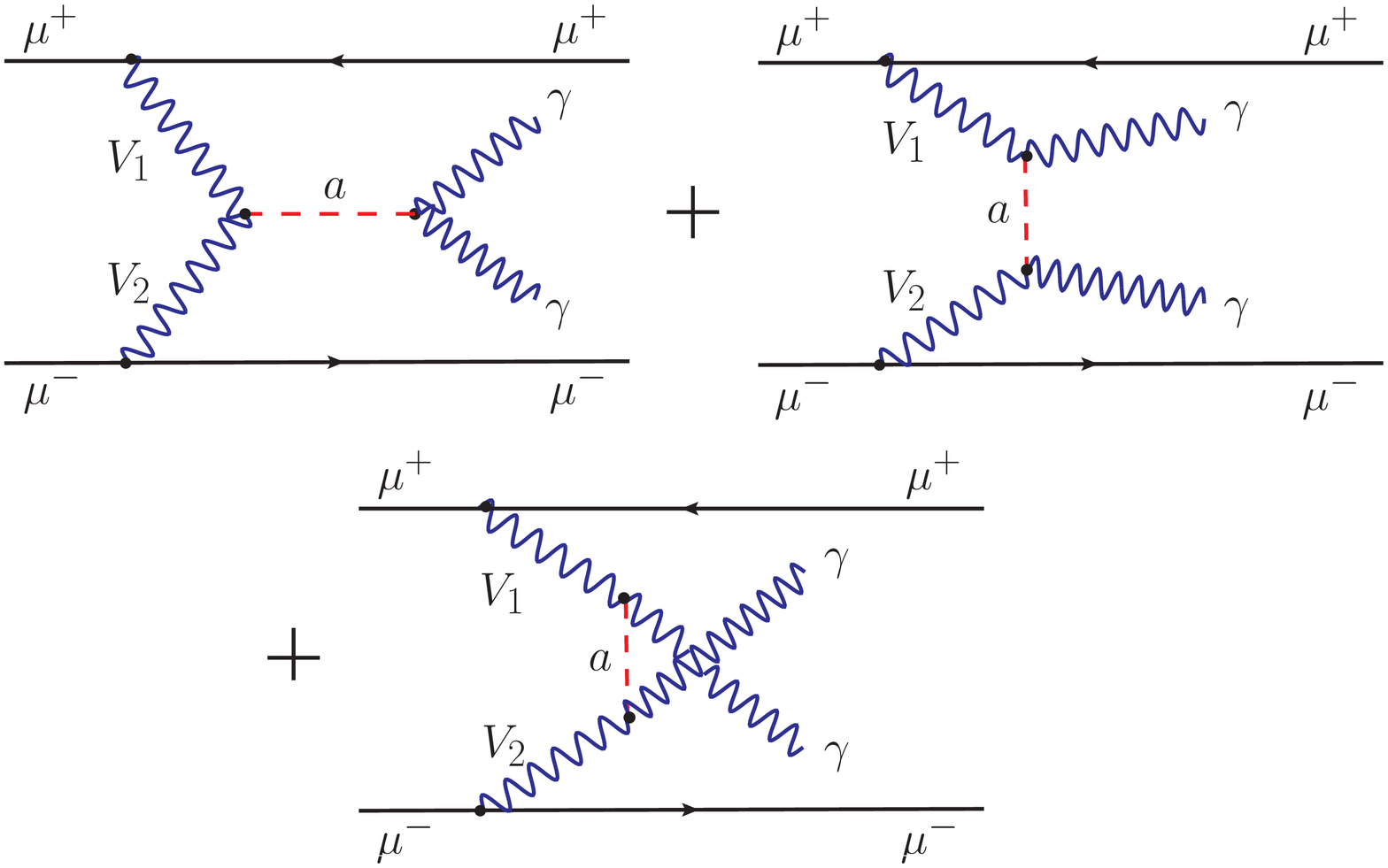}
\caption{The Feynman diagrams describing virtual production of the
axion-like particle $a$ in the $\mu^+\mu^-$ collision via vector
boson fusion.}
\label{fig:mu-VBF-mu}
\end{center}
\end{figure}

We consider the process shown in \ref{fig:mu-VBF-mu}.
%
In the equivalent photon approximation (EPA)
\cite{Weizsacker:1934}-\cite{Carimalo:1979}, the photon has the
following leading logarithmic approximation spectrum
\cite{Budnev:1975}
\begin{equation}\label{photon_spectrum}
f_{\gamma/\mu^\pm}(x, Q^2) = \frac{\alpha}{2\pi} \frac{1 + (1 -
x)^2}{x} \ln\frac{Q^2}{m_\mu^2} \;,
\end{equation}
where $x = E_\gamma/E_\mu$ is the ratio of the photon energy
$E_\gamma$ and energy of the incoming muon $E_\mu$, $m_\mu$ is the
muon mass. To examine the collisions of massive vector bosons
($W^\pm$ and $Z$), the effective $W$ approximation (EWA) is applied
\cite{Dawson:1985,Kane:1984} which allows to treat massive vector
bosons as partons inside the colliding beams (see also
\cite{Cahn:1984}-\cite{Ruiz:2021}). In this scheme the $Z$ boson has
different distributions for its transverse ($T$) and longitudinal
($L$) polarizations. The leading order distributions of the $Z$
boson inside the colliding muon are the following
\cite{Lindfors:1987,Ruiz:2021}
\begin{align}\label{Z_spectrum}
f_{Z_T/\mu^\pm}(x, Q^2) &= \frac{\alpha_Z^\pm}{2\pi} \frac{1 + (1 -
x)^2}{x}
\ln\frac{Q^2}{m_Z^2} \;, \nonumber \\
f_{Z_L/\mu^\pm}(x, Q^2) &= \frac{\alpha_Z^\pm}{\pi} \frac{(1 -
x)}{x} \;,
\end{align}
where $x = E_Z/E_\mu$ is the fraction of the muon energy held by
$Z$, and
\begin{equation}\label{alpha_Z}
\alpha_Z^\pm = \frac{\alpha}{(\cos\theta_W \sin\theta_W)^2} \left[
(g_V^\pm)^2 + (g_A^\pm)^2 \right] ,
\end{equation}
with $g_V^{\pm} = -1/4 \mp \sin^2\theta_W$, $g_A^{\pm} = 1/4$.

Remember that we examine the \emph{exclusive} process $\mu^+\mu^-
\rightarrow \mu^+\mu^- + \gamma\gamma$. Should one study the
\emph{inclusive} process $\mu^+\mu^- \rightarrow \gamma\gamma + X$,
where $X$ is an unspecified remnant, he has to use electroweak
parton distribution functions (see \cite{Han:2021_3} and references
therein).

The cross section of our process $\mu^{-}\mu^{+} \rightarrow \mu^{-}
V_1 V_2 \mu^{+} \rightarrow  \mu^{-}\gamma\gamma \mu^{+}$ is defined
by the formula
\begin{equation}\label{cs}
d\sigma = \int\limits_{\tau_{\min}}^{\tau_{\max}} \!\!d\tau
\!\!\!\int\limits_{x_{\min}}^{x_{\max}} \!\!\frac{dx}{x}
\,\sum_{V_1, V_2 = \gamma, Z_T,Z_L} \!\!f_{V_1/\mu^+}(x, Q^2)
f_{V_2/\mu^-}(\tau/x, Q^2) \,d\hat{\sigma} (V_1V_2\rightarrow
\gamma\gamma) \;.
\end{equation}
Here
\begin{equation}\label{y_z_limits}
x_{\max} = 1 - \frac{m_\mu}{E_\mu} \;, \ \tau_{\max} = \left( 1 -
\frac{m_\mu}{E_\mu} \right)^{\!2} , \ x_{\min} = \tau/x_{\max} \;, \
\tau_{\min} = \frac{p_\bot^2}{E_\mu^2} \;,
\end{equation}
and $p_\bot$ is the transverse momenta of the outgoing photons. The
boson distributions inside the muon beam, $f_{\gamma/{\mu^\pm}}(x,
Q^2)$, $f_{Z_T/{\mu^\pm}}(x, Q^2)$, and $f_{Z_L/{\mu^\pm}}(x, Q^2)$
are given by eqs.~\eqref{photon_spectrum} and \eqref{Z_spectrum},
respectively, and the subprocess cross section $d\hat{\sigma}
(V_1V_2\rightarrow\gamma\gamma)$ is defined by
eq.~\eqref{subprocess_cs}. We take $Q^2 = \hat{s}$, where
$\sqrt{\hat{s}} = 2E_\mu \sqrt{\tau}$ is the invariant energy of the
VBF subprocess $V_1V_2\rightarrow\gamma\gamma$.

\subsection{Numerical analysis} 

\begin{figure}[htb]
\begin{center}
\includegraphics[scale=0.5]{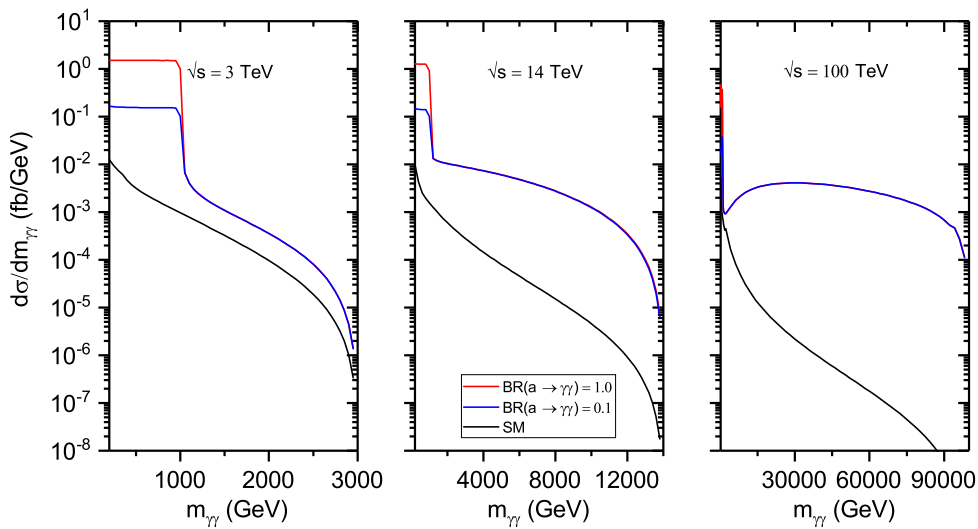}
\caption{The differential cross sections for the  $\mu^+\mu^-
\rightarrow \mu^+ \gamma\gamma \mu^-$ scattering at the future muon
collider versus diphoton invariant mass $m_{\gamma\gamma}$. The
curves correspond to the ALP mass $m_a = 1$ TeV and ALP-gauge boson
coupling $f_a = 10$ TeV.} \label{fig:DMALP}
\end{center}
\end{figure}

\begin{figure}[htb]
\begin{center}
\includegraphics[scale=0.5]{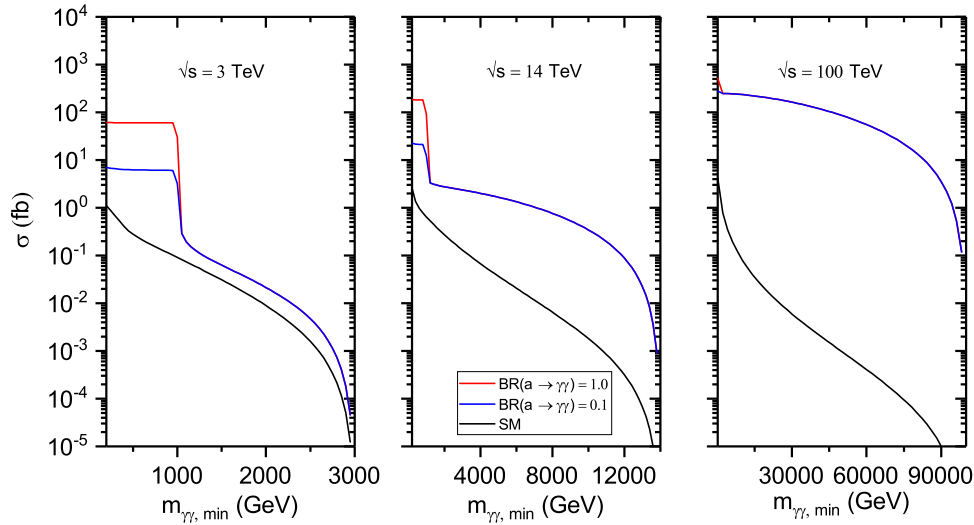}
\caption{The total cross sections $\sigma(m_{\gamma\gamma} >
m_{\gamma\gamma,\min})$ for the $\mu^+\mu^- \rightarrow \mu^+
\gamma\gamma \mu^-$ scattering at the future muon collider versus
minimal value of the diphoton invariant mass $m_{\gamma\gamma}$. The
curves correspond to the ALP mass $m_a = 1$ TeV and ALP-gauge boson
coupling $f_a = 10$ TeV.} \label{fig:MCUTALP}
\end{center}
\end{figure}

The results of our calculations of the differential cross section
for the $\mu^+\mu^- \rightarrow \mu^+ \gamma\gamma \mu^-$ collision
at the future muon collider are presented in Fig.~\ref{fig:DMALP}.
The collision energies $\sqrt{s}$ of 3 TeV, 14 TeV and 100 TeV, and
two values of the ALP branching are addressed. For comparison, the
SM predictions are shown. As one can see, a discrepancy between the
cross section and its SM part rises significantly as the collision
energy grows. The same tendency takes place for the total cross
section $\sigma(m_{\gamma\gamma} > m_{\gamma\gamma,\min})$, where
$m_{\gamma\gamma,\min}$ is the minimal invariant mass of the final
photons $m_{\gamma\gamma}$, see Fig.~\ref{fig:MCUTALP}. Moreover,
the total cross section becomes larger with the increase of $s$ in
the whole region of $m_{\gamma\gamma,\min}$. We have used the cut on
the rapidity and transverse momentum of the outgoing photons,
$|\eta| < 2.5$ and $p_t > 30$ GeV respectively in all calculations.

\begin{figure}[htb]
\begin{center}
\includegraphics[scale=0.65]{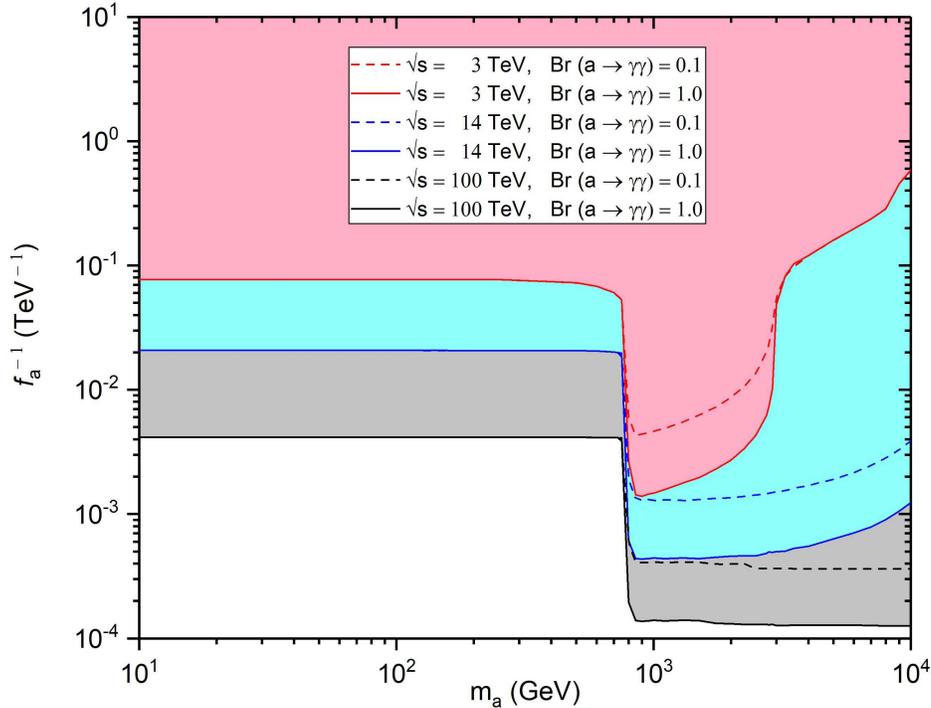} \\
\caption{95\% C.L.exclusion regions for the ALP-gauge boson coupling
$f_a$ coming from the $\mu^+\mu^- \rightarrow \mu^+ \gamma\gamma
\mu^-$ scattering at the future muon collider. The curves are
obtained with the use of the cut on diphoton invariant mass,
$m_{\gamma\gamma} > 800$ GeV.} \label{fig:ALP_EXCLUSION}
\end{center}
\end{figure}

To derive the exclusion region, we apply the following formula for
the statistical significance $SS$ \cite{SS}
\begin{equation}\label{SS_def}
SS = \sqrt{2[(S - B\,\ln(1 + S/B)]} \;,
\end{equation}
where $S$ is the number of signal events and $B$ is the number of
background (SM) events. We define the regions $SS \leqslant 1.645$
as the regions that can be excluded at the 95\% C.L. To reduce the
SM background, we used the cut $m_{\gamma\gamma} > 800$ GeV. The
results are shown in Fig.~\ref{fig:ALP_EXCLUSION}. Following
\cite{MC_group:1999} (see also \cite{Han:2021}), we consider the
integrated luminosities of 1 ab$^{-1}$, 20 ab$^{-1}$, and 1000
ab$^{-1}$ for the muon collider energies of 3 TeV, 14 TeV, and 100
TeV, respectively.

As one can see in Fig.~\ref{fig:ALP_EXCLUSION}, for $\sqrt{s} = 3$
TeV the best sensitivity region is limited to a rather sharp region
800 GeV -- 2 TeV. In this region the ALP term dominates, while
outside it the contributions from the ALP and SM terms are
comparable and they partially cancel each other. For $\sqrt{s} = 14$
TeV and $\sqrt{s} = 100$ TeV the ALP contributions dominate in wider
regions of the ALP mass ($\geq 800$ GeV). A similar effect was shown
to take place for the ALP production in the LBL scattering
\cite{I_K:2020} (for details, see Appendix~A in \cite{I_K:2020}).
For comparison, in Fig.~\ref{fig:CBSSPOL_all_f} we present
previously obtained $95\%$ C.L. exclusion region for the ALP-photon
coupling coming from the polarized LBL scattering at the 3 TeV CLIC
\cite{I_K:2020}. Other current exclusion regions for this coupling
are also shown \cite{Beldenegro:2018}. As seen from the
Figs.~\ref{fig:ALP_EXCLUSION} and \ref{fig:CBSSPOL_all_f}, the
excluded areas that we have found from studying diphoton production
at the muon collider extends to wider regions, especially for
$\sqrt{s}=14$ TeV and $\sqrt{s}=100$ TeV.

\begin{figure}[htb]
\begin{center}
\includegraphics[scale=0.2]{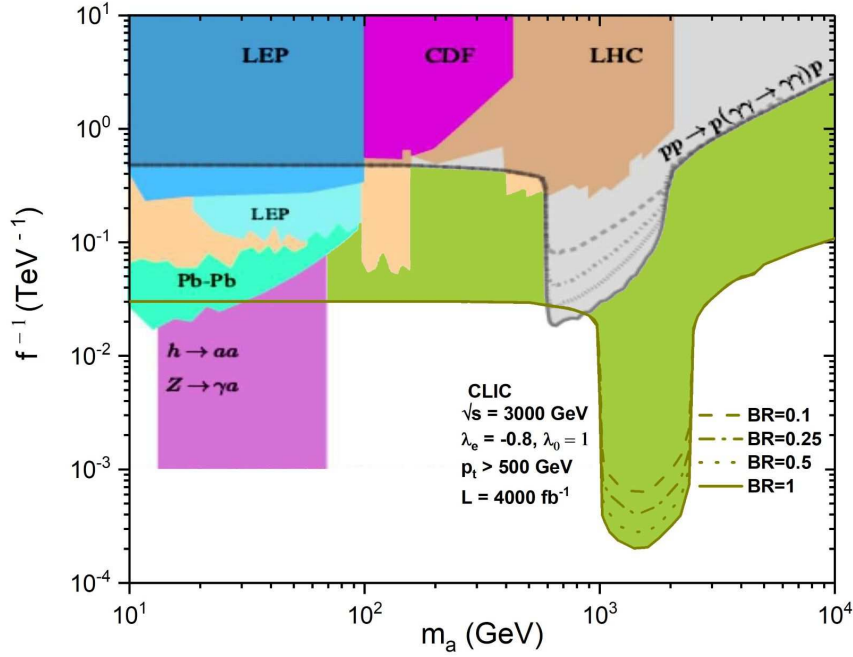}
\caption{Our previous $95\%$ C.L. exclusion region for the
ALP-photon coupling in the polarized light-by-light scattering at
the 3 TeV CLIC induced by ALPs (green area) \protect\cite{I_K:2020}
in comparison with other current exclusion regions
\protect\cite{Beldenegro:2018}.}
\label{fig:CBSSPOL_all_f}
\end{center}
\end{figure}

\section{Unitarity constraints on ALP coupling} %
\label{sec:unit_const}

Let us study bounds imposed by partial-wave unitarity. The
partial-wave expansion of the helicity amplitude in the
center-of-mass system was derived in \cite{Jacob:2000}. It looks
like
\begin{align}\label{helicity_ampl_expansion}
M_{\lambda_1\lambda_2\lambda_3\lambda_4}(s, \theta, \varphi) &=
16\pi \sum_J (2J + 1) \sqrt{(1 + \delta_{\lambda_1\lambda_2})(1 +
\delta_{\lambda_3\lambda_4})}
\nonumber \\
&\times \,e^{i(\lambda - \mu)\phi} \,d^J_{\lambda\mu}(\theta)
\,T^J_{\lambda_1\lambda_2\lambda_3\lambda_4}(s) \;,
\end{align}
where $\lambda = \lambda_1 - \lambda_2$, $\mu = \lambda_3 -
\lambda_4$, $\theta(\phi)$ is the polar (azimuth) scattering angle,
and $d^J_{\lambda\mu}(\theta)$ is the Wigner (small) $d$-function
\cite{Wigner}. Relevant formulas for the $d$-functions can be found
in \cite{I_K:2021_3}. If we choose the plane $(x - z)$ as a
scattering plane, then $\phi = 0$ in
\eqref{helicity_ampl_expansion}. Parity conservation means that
\begin{equation}\label{parity_conservation}
T^J_{\lambda_1\lambda_2\lambda_3\lambda_4}(s) = (-1)^{\lambda_1 -
\lambda_2 - \lambda_3 + \lambda_4}
\,T^J_{-\lambda_1-\lambda_2-\lambda_3-\lambda_4}(s) \;.
\end{equation}
Partial-wave unitarity in the limit $s \gg (m_1 + m_2)^2$ requires
that
\begin{equation}\label{parity_wave_unitarity}
\left| T^J_{\lambda_1\lambda_2\lambda_3\lambda_4}(s) \right| \leq 1
\;.
\end{equation}
Using orthogonality of the $d$-functions,
\begin{equation}\label{d_func_orthogonality}
\int\limits_{-1}^1 d^{J}_{\lambda\lambda'}(z)
\,d^{J'}_{\lambda\lambda'}(z) \,dz = \frac{2}{2J + 1} \,\delta_{JJ'}
\;,
\end{equation}
we find from \eqref{helicity_ampl_expansion} that the partial-wave
amplitude is defined as
\begin{align}\label{parity wave_func}
T^J_{\lambda_1\lambda_2\lambda_3\lambda_4}(s) &= \frac{1}{32\pi}
\frac{1}{\sqrt{(1 + \delta_{\lambda_1\lambda_2})(1 +
\delta_{\lambda_3\lambda_4})}} \int\limits_{-1}^1 \!\!
M_{\lambda_1\lambda_2\lambda_3\lambda_4}(s, z)
\,d^{J}_{\lambda\mu}(z) \,dz \;.
\end{align}
Here and in what follows, $z = \cos\theta$. The helicity amplitudes
$M_{\lambda_1\lambda_2\lambda_3\lambda_4}$ are given in Appendix~A.

The $d$-functions obey, inter alia, the relation
$d^J_{\lambda\mu}(-z) = (-1)^{J - \lambda}d^J_{\mu -\lambda}(z)$. In
particular, we have ($J \geq 0$)
\begin{equation}\label{d-function_00}
d^J_{00}(z) = P_J(z) \;,
\end{equation}
$P_J(z)$ being the Legendre polynomial, and \cite{I_K:2021_3}
\begin{align}
d^J_{2-2}(z) &= (-1)^J \left( \frac{1 - z}{2}\right)^{\!2} {}_2F_1
\!\left( 2-J, J+3; 1; \frac{1+z}{2} \right) , \label{d-function_2-2}
\\
d^J_{22}(z) &= \left( \frac{1 + z}{2}\right)^{\!2} {}_2F_1 \!\left(
2-J, J+3; 1; \frac{1-z}{2} \right) , \label{d-function_22}
\end{align}
where $_2F_1(a,b;c;x)$ is the hypergeometric function
\cite{Bateman_vol_1}, and $J \geq 2$.

\textbf{1.} Consider the helicity amplitude
$M_{++++}^{\gamma\gamma}$ \eqref{M++++gamma}. Then $\lambda_1 =
\lambda_2 = \lambda_3 = \lambda_4 = 1$, and $\lambda = \mu = 0$.
Since $s, m_a^2 \gg m_a \Gamma_a$, we can write
\begin{equation}\label{M++++gamma_u}
M_{++++}^{\gamma\gamma}(s,z) = -\frac{4}{f_a^2}\,\frac{s^2}{s -
m_a^2} \;.
\end{equation}
The partial-wave amplitude with $J=0$ is the only non-zero
amplitude, since
\begin{equation}\label{TJ++++gamma}
T^J_{++++}(s) = -\frac{1}{16\pi f_a^2}\,\frac{s^2}{s - m_a^2}
\int\limits_{-1}^1 P_J(z) \,dz = -\frac{1}{8\pi f_a^2}\,\frac{s^2}{s
- m_a^2} \,\delta_{J0} \;.
\end{equation}
Then we obtain from \eqref{parity_wave_unitarity},
\eqref{TJ++++gamma} the unitarity bound on the the ALP-gauge boson
coupling coupling
\begin{equation}\label{T0++++gamma_bound}
f_a^2 \geq  \frac{1}{8\pi}\,\frac{s}{|1 - \varepsilon|} \;,
\end{equation}
where $\varepsilon = m_a^2/s$.

\textbf{2.} For the helicity amplitude $M_{++--}^{\gamma\gamma}$
\eqref{M++--gamma} we have $\lambda_1 = \lambda_2 = 1$, $\lambda_3 =
\lambda_4 = -1$, and, consequently, $\lambda = 0$, $\mu = 0$. It
looks like \eqref{M++--gamma}
\begin{equation}\label{M++--gamma_u}
M_{++--}^{\gamma\gamma}(s,z) = \frac{2s}{f_a^2} \left[ \frac{2}{1 -
\varepsilon} - \frac{(1 - z)^2}{1 - z + 2\varepsilon} - \frac{(1 +
z)^2}{1 + z + 2\varepsilon} \right] .
\end{equation}
As a result, we obtain
\begin{align}\label{TJ++--gamma_1}
T^J_{++--}(s) &= \frac{s}{32\pi f_a^2} \left[ \frac{2}{1 -
\varepsilon} \int\limits_{-1}^1 \!P_J(z) \,dz  - \left( 1 + (-1)^J
\right) \int\limits_{-1}^1 \!\frac{(1 - z)^2}{1 - z + 2\varepsilon}
P_J(z)
\,dz \right] \nonumber \\
&= \frac{s}{8\pi f_a^2} \left[ \frac{\varepsilon (3 -
2\varepsilon)}{1 - \varepsilon} \,\delta_{J0} - \left( 1 + (-1)^J
\right)\varepsilon^2 Q_J(1 + 2\varepsilon) \right] ,
\end{align}
where $Q_J(x)$ is the Legendre function of the second kind
\cite{Bateman_vol_2}. If $x > 1$, $Q_J(x)$ is a real strictly
decreasing function of $J$, and it decreases exponentially as $J
\rightarrow \infty$. Note that $Q_J(1 + 2\varepsilon) \simeq -(\ln
\varepsilon)/2$ for $\varepsilon \ll 1$, $J\geq 0$. The term with
$J=0$ is a leading one,
\begin{equation}\label{TJ++--gamma_2}
T^0_{++--}(s) = \frac{s}{8\pi f_a^2} \,\varepsilon \!\left[ \frac{3
- 2\varepsilon}{1 - \varepsilon} - \varepsilon \ln
\frac{1+\varepsilon}{\varepsilon} \right] .
\end{equation}
It results in the following unitarity bound
\begin{equation}\label{T0+--+gamma_bound}
f_a^2 \geq \frac{s}{8\pi} \,\varepsilon \!\left| \frac{3 -
2\varepsilon}{1 - \varepsilon} - \varepsilon \ln
\frac{1+\varepsilon}{\varepsilon} \right| .
\end{equation}

\textbf{3.} Now consider the helicity amplitude
$M_{+--+}^{\gamma\gamma}$ \eqref{M+--+gamma}. Then $\lambda_1 =
\lambda_4 = 1$, $\lambda_2 = \lambda_3 = -1$, and $\lambda = 2$,
$\mu = -2$. The helicity amplitude is given by
eq.~\eqref{M+--+gamma},
\begin{equation}\label{M+--+gamma_u}
M_{+--+}^{\gamma\gamma}(s,z) = \frac{2}{f_a^2}\,\frac{s(1 - z)^2}{1
- z + 2\varepsilon} \;.
\end{equation}
Then we get from \eqref{parity wave_func}, \eqref{d-function_2-2},
\eqref{M+--+gamma_u}
\begin{align}\label{TJ+--+gamma_2}
T^J_{+--+}(s) &= (-1)^J \,\frac{s}{64\pi f_a^2} \int\limits_{-1}^1
\!\frac{(1 - z)^4}{1 - z + 2\varepsilon} \,\,{}_2F_1 \!\left( 2-J,
J+3; 1; \frac{1+z}{2} \right) dz
\nonumber \\
&= (-1)^J \,\frac{s}{4\pi f_a^2} \,I(J,\varepsilon) \;,
\end{align}
where the notation
\begin{equation}\label{I}
I(J,\varepsilon) = \int\limits_0^1 \!\frac{x^4}{x + \varepsilon}
\,\,{}_2F_1(2-J, J+3; 1; 1 - x) \,dx
\end{equation}
is introduced. Using formula 2.21.1.26 in \cite{Prudnikov_v3}, we
obtain a sequence of two equalities (recall that $J \geq 2$)
\begin{align}\label{hiper_geom_int}
I(J,\varepsilon) &= \frac{\Gamma(5)}{\varepsilon
\,\Gamma(3-J)\Gamma(J+4)} \, {}_3F_2\!\left( 1; 1; 5; 3-J; J+4;
-\frac{1}{\varepsilon} \right)
\nonumber \\
&=(-1)^J \,\frac{\Gamma(J-1)\Gamma(J+3)}{\varepsilon^{J-1}
\,\Gamma(2J + 2)} \,{}_2F_1\!\left( J-1, J+3; 2J + 2;
-\frac{1}{\varepsilon} \right) ,
\end{align}
where $\Gamma(x)$ denotes the gamma function \cite{Bateman_vol_1}.
In \eqref{hiper_geom_int} we have reduced a generalized
hypergeometric function ${}_3F_2(a,b,c; d,e;x)$ to a traditional
hypergeometric function. With a help of equation 2.10(6) in
\cite{Bateman_vol_1} we find the final analytic expression for
$I(J,\varepsilon)$,
\begin{align}\label{hiper_geom_relation_2}
I(J,\varepsilon) &= (-1)^J (1+\varepsilon)^{1-J}
\,\frac{\Gamma(J-1)\Gamma(J+3)}{\Gamma(2J + 2)} \,
\nonumber \\
&\times {}_2F_1\!\left( J-1, J-1; 2J + 2; \frac{1}{1+\varepsilon}
\right) .
\end{align}
Using integral representation for the hypergeometric function (see
formula 2.12(1) in \cite{Bateman_vol_1}), one can show that for
$\varepsilon > 0$ the right-hand side of
eq.~\eqref{hiper_geom_relation_2} is a strictly decreasing function
of $J$. Moreover, it falls off exponentially at large $J$. Thus, the
most stringent unitarity bound comes from the partial-wave amplitude
with $J=2$ that looks like
\begin{align}\label{T2+--+gamma}
T^2_{+--+}(s) &= \frac{s}{20\pi f_a^2} \frac{1}{1+\varepsilon}
\,{}_2F_1\!\left( 1, 1; 6; \frac{1}{1+\varepsilon} \right)
\nonumber \\
&= \frac{s}{16\pi f_a^2} \left[ 1 - \frac{4\varepsilon}{3} +
2\varepsilon^2 - 4\varepsilon^3 + 4\varepsilon^4 \ln
\frac{1+\varepsilon}{\varepsilon} \right] .
\end{align}
As a result, we come to the unitarity bound
\begin{equation}\label{T2+--+gamma_bound}
f_a^2 \geq \frac{s}{16\pi} \left| 1 - \frac{4\varepsilon}{3} +
2\varepsilon^2 - 4\varepsilon^3 + 4\varepsilon^4 \ln
\frac{1+\varepsilon}{\varepsilon} \right| .
\end{equation}
Note that the right-hand side of this equation does not exceed
$s/(16\pi)$.

The analogous examination of the amplitude
$M_{+-+-}^{\gamma\gamma}$, using eqs.~\eqref{M+-+-gamma} and
\eqref{d-function_22}, results in just the same bound
\eqref{T2+--+gamma_bound}.

The unitarity constraints for the amplitudes
$M_{\lambda_1\lambda_2\lambda_3\lambda_4}^{Z\gamma}$ and
$M_{\lambda_1\lambda_2\lambda_3\lambda_4}^{ZZ}$ differ from the
above presented bounds for
$M_{\lambda_1\lambda_2\lambda_3\lambda_4}^{\gamma\gamma}$ (with the
same helicities) by the factors $2s_w/c_w \simeq 1.1$ and
$s_w^2/c_w^2 \simeq 0.3$, respectively, neglecting small corrections
of $\mathrm{O}(m_Z/\sqrt{s})$ or $\mathrm{O}(m_Z^2/s)$. In
particular, imposing unitarity constraint on the amplitude
$M_{++++}^{Z\gamma}$, we get the lower bound (up to small
corrections $\mathrm{O}(m_Z^2/s)$)
\begin{equation}\label{T0++++Z_bound}
f_a^2 \geq \frac{1}{4\pi}\frac{s_w}{c_w}\,\frac{s}{|1 -
\varepsilon|} \;.
\end{equation}
This constraint is slightly stronger than \eqref{T0++++gamma_bound}.
Unitarity bounds for the most of the other helicity amplitudes with
the $Z$ boson(s) are suppressed by small factors $m_Z/\sqrt{s}$ or
$m_Z^2/s$.

The constraint \eqref{T0++++Z_bound} appears to be the strongest
unitary bound. We find from it that for $m_a = 800$ GeV the inverse
ALP coupling $f_a^{-1}$ must be smaller than 1.54 TeV$^{-1}$, 0.34
TeV$^{-1}$, and 0.048 TeV$^{-1}$, for the collision energies of 3
TeV, 14 TeV, and 100 TeV, respectively. Correspondingly, for $m_a =
10$ TeV the inverse coupling should be less than 5.01 TeV$^{-1}$,
0.24 TeV$^{-1}$, and 0.048 TeV$^{-1}$. By comparing these
constraints with the curves presented in
Fig.~\ref{fig:ALP_EXCLUSION}, we conclude that the unitarity is not
violated in the region of the ALP coupling $f_a$ studied in the
present paper.

\section{Conclusions} %

We have examined the possibility to search for heavy axion-like
particles in the $\mu^+\mu^- \rightarrow \mu^+ \gamma\gamma \mu^-$
scattering at the future muon collider. The studies are presented
for the collision energies of 3 TeV, 14 TeV, and 100 TeV and
integrated luminosities of 1 ab$^{-1}$, 20 ab$^{-1}$, and 1000
ab$^{-1}$, respectively. We have obtained the explicit expressions
for the helicity amplitudes for the $Z\gamma \rightarrow
\gamma\gamma$ and $ZZ \rightarrow \gamma\gamma$ collisions. Using
these amplitudes (as well as known helicity amplitudes for the
$\gamma\gamma \rightarrow \gamma\gamma$ collision), the differential
cross sections versus invariant mass of the final photons and total
cross section versus minimal diphoton invariant mass are calculated.
As a result, the 95\% C.L.exclusion regions for the ALP-gauge boson
coupling coming from the $\mu^+\mu^- \rightarrow \mu^+ \gamma\gamma
\mu^-$ scattering at the high energy muon collider are obtained. The
excluded areas extend to wider regions in comparison to the region
obtained previously for the polarized light-by-light scattering at
the 3 TeV CLIC. Our constraints are also much stronger than the
current experimental bounds presented in
Fig.~\ref{fig:CBSSPOL_all_f}. The partial-wave unitarity bounds on
the ALP-gauge boson coupling are estimated. We have shown that the
unitarity is not violated in the region of the ALP coupling which
has been studied in our paper. We can conclude that the future muon
collider has a great physical potential in searching for axion-like
particle couplings to the SM gauge bosons.


\setcounter{equation}{0}
\renewcommand{\theequation}{A.\arabic{equation}}

\section{Appendix A. Helicity amplitudes} %

\subsection{$\gamma\gamma \rightarrow \gamma\gamma$ scattering} %

The Mandelstam variables for the $\gamma\gamma \rightarrow
\gamma\gamma$ collision satisfy the relation $s + t + u = 0$, and we
get
\begin{equation}\label{cos_sin_gamma}
\cos\theta = \frac{u - t}{u + t} \;, \quad \sin\theta =
-\frac{2\sqrt{tu}}{t + u} \;,
\end{equation}
\begin{equation}\label{t_u_gamma}
t = - \frac{s}{2}\,(1 - \cos\theta) \;, \quad u = - \frac{s}{2}\,(1
+ \cos\theta) \;.
\end{equation}
The helicity amplitudes of the LBL scattering are known to be
\cite{Beldenegro:2018}

\begin{equation}\label{M++++gamma}
M_{++++}^{\gamma\gamma} = -\frac{4}{f_a^2}\,\frac{s^2}{s - m_a^2 +
im_a \Gamma_a} \;,
\end{equation}

\begin{equation}\label{M+++-gamma}
M_{+++-}^{\gamma\gamma} = 0 \;,
\end{equation}

\begin{equation}\label{M++-+gamma}
M_{++-+}^{\gamma\gamma} = 0 \;,
\end{equation}

\begin{equation}\label{M+-++gamma}
M_{+-++}^{\gamma\gamma} 0 \;,
\end{equation}

\begin{align}\label{M++--gamma}
M_{++--}^{\gamma\gamma} &= \frac{4}{f_a^2}\,\frac{s^2}{s - m_a^2 +
im_a \Gamma_a} \nonumber
\\
&+ \frac{s^2}{f_a^2} \left[ \frac{(1 - \cos\theta)^2}{t - m_a^2 +
im_a \Gamma_a} + \frac{(1 + \cos\theta)^2}{u - m_a^2 + im_a
\Gamma_a} \right] ,
\end{align}

\begin{equation}\label{M+--+gamma}
M_{+--+}^{\gamma\gamma} = -\frac{1}{f_a^2}\,\frac{s^2(1 -
\cos\theta)^2}{t - m_a^2 + im_a \Gamma_a} \;,
\end{equation}

\begin{equation}\label{M+-+-gamma}
M_{+-+-}^{\gamma\gamma} = -\frac{1}{f_a^2}\,\frac{s^2(1 +
\cos\theta)^2}{u - m_a^2 + im_a \Gamma_a} \;,
\end{equation}

\begin{equation}\label{M+---gamma}
M_{+---}^{\gamma\gamma} = 0 \;.
\end{equation}
Other helicity amplitudes
$M_{\lambda_1\lambda_2\lambda_3\lambda_4}^{\gamma\gamma}$ can be
obtained by the $P$-parity relation
\begin{equation}\label{parity_relation_gamma}
M_{\lambda_1\lambda_2\lambda_3\lambda_4}^{\gamma\gamma} =
M_{-\lambda_1-\lambda_2-\lambda_3-\lambda_4}^{\gamma\gamma} \;.
\end{equation}

\subsection{$Z\gamma \rightarrow \gamma\gamma$ scattering} %

The Mandelstam variables for this process obey the relation $s + t +
u = m_Z^2$, variables $\cos\theta$, $\sin\theta$ are given by
eq.~\eqref{cos_sin_gamma}, and
\begin{equation}\label{t_u_Z}
t = - \frac{s - m_Z^2}{2} \,(1 - \cos\theta) \;, \quad u = - \frac{s
- m_Z^2}{2}\,(1 + \cos\theta) \;.
\end{equation}
Our calculations result in the following analytic expressions for
the helicity amplitudes of the $Z\gamma \rightarrow \gamma\gamma$
process:
\begin{equation}\label{M++++Z}
M_{++++}^{Z\gamma} = \frac{8s_w}{c_w}\frac{1}{f_a^2}\,\frac{s(s -
m_Z^2)}{s - m_a^2 + im_a \Gamma_a} \;,
\end{equation}

\begin{equation}\label{M+++-Z}
M_{+++-}^{Z\gamma} = \frac{2s_w}{c_w}\frac{1}{f_a^2}\,\frac{m_Z^2(s
- m_Z^2)(\sin\theta)^2}{t - m_a^2 + im_a \Gamma_a} \;,
\end{equation}

\begin{equation}\label{M++-+Z}
M_{++-+}^{Z\gamma} = \frac{2s_w}{c_w}\frac{1}{f_a^2}\,\frac{m_Z^2(s
- m_Z^2) (\sin\theta)^2}{u - m_a^2 + im_a \Gamma_a} \;,
\end{equation}

\begin{align}\label{M+-++Z}
M_{+-++}^{Z\gamma} &= -\frac{2s_w}{c_w}\frac{1}{f_a^2}\,m_Z^2(s -
m_Z^2)(\sin\theta)^2
\nonumber \\
&\times \left[ \frac{1}{t - m_a^2 + im_a \Gamma_a} + \frac{1}{u -
m_a^2 + im_a \Gamma_a} \right] ,
\end{align}

\begin{align}\label{M++--Z}
M_{++--}^{Z\gamma} &= -\frac{8s_w}{c_w}\frac{1}{f_a^2}\,\frac{s(s -
m_Z^2)}{s - m_a^2 + im_a \Gamma_a}
-\frac{2s_w}{c_w}\frac{1}{f_a^2}\,s(s - m_Z^2) \nonumber
\\
&\times \left[ \frac{(1 - \cos\theta)^2}{t - m_a^2 + im_a \Gamma_a}
+ \frac{(1 + \cos\theta)^2}{u - m_a^2 + im_a \Gamma_a} \right] ,
\end{align}

\begin{equation}\label{M+--+Z}
M_{+--+}^{Z\gamma} = \frac{2s_w}{c_w}\frac{1}{f_a^2}\,\frac{s(s -
m_Z^2)(1 - \cos\theta)^2}{t - m_a^2 + im_a \Gamma_a} \;,
\end{equation}

\begin{equation}\label{M+-+-Z}
M_{+-+-}^{Z\gamma} = \frac{2s_w}{c_w}\frac{1}{f_a^2}\,\frac{s(s -
m_Z^2)(1 + \cos\theta)^2}{u - m_a^2 + im_a \Gamma_a} \;,
\end{equation}

\begin{equation}\label{M+---Z}
M_{+---}^{Z\gamma} = 0 \;,
\end{equation}

\begin{equation}\label{M0+++Z}
M_{0+++}^{Z\gamma} = 0 \;,
\end{equation}

\begin{equation}\label{M0++-Z}
M_{0++-}^{Z\gamma} =  \frac{4i}{\sqrt{2}}
\frac{s_w}{c_w}\frac{1}{f_a^2}\,\frac{m_Z \sqrt{s}(s - m_Z^2)(1 -
\cos\theta)\sin\theta}{t - m_a^2 + im_a \Gamma_a} \;,
\end{equation}

\begin{equation}\label{M0+-+Z}
M_{0+-+}^{Z\gamma} = -\frac{4i}{\sqrt{2}}
\frac{s_w}{c_w}\frac{1}{f_a^2} \,\frac{m_Z \sqrt{s}(s - m_Z^2)(1 +
\cos\theta)\sin\theta}{u - m_a^2 + im_a \Gamma_a} \;,
\end{equation}

\begin{align}\label{M0-++Z}
M_{0-++}^{Z\gamma} &=
\frac{4i}{\sqrt{2}}\frac{s_w}{c_w}\frac{1}{f_a^2} \,
m_Z \sqrt{s}(s - m_Z^2) \sin\theta \nonumber\\
&\times \left[ \frac{(1 - \cos\theta)}{t - m_a^2 + im_a \Gamma_a} -
\frac{(1 + \cos\theta)}{u - m_a^2 + im_a \Gamma_a} \right] ,
\end{align}

\begin{align}\label{M0+--Z}
M_{0+--}^{Z\gamma} &=
-\frac{4i}{\sqrt{2}}\frac{s_w}{c_w}\frac{1}{f_a^2} \,
m_Z \sqrt{s}(s - m_Z^2)\sin\theta \nonumber\\
&\times \left[ \frac{(1 - \cos\theta)}{t - m_a^2 + im_a \Gamma_a} -
\frac{(1 + \cos\theta)}{u - m_a^2 + im_a \Gamma_a} \right] ,
\end{align}

\begin{equation}\label{M0--+Z}
M_{0--+}^{Z\gamma} = \frac{4i}{\sqrt{2}}
\frac{s_w}{c_w}\frac{1}{f_a^2}\,\frac{m_Z \sqrt{s}(s - m_Z^2)(1 -
\cos\theta)\sin\theta}{t - m_a^2 + im_a \Gamma_a} \;.
\end{equation}
Other amplitudes
$M_{\lambda_1\lambda_2\lambda_3\lambda_4}^{Z\gamma}$ can be obtained
by relation \cite{Gounaris:1999_3}
\begin{equation}\label{parity_relation_Z}
M_{\lambda_1\lambda_2\lambda_3\lambda_4}^{Z\gamma} =
(-1)^{1-\lambda_1}
M_{-\lambda_1-\lambda_2-\lambda_3-\lambda_4}^{Z\gamma} \;,
\end{equation}
where $\lambda_1$ is a helicity of the $Z$ boson.

\subsection{$ZZ \rightarrow \gamma\gamma$ scattering} %

The Mandelstam variables obey the relation $s + t + u = 2m_Z^2$, and
we obtain
\begin{equation}\label{cos_sin_ZZ}
\cos\theta = \frac{t - u}{\sqrt{(t + u)^2 - 4m_Z^4}} \;, \quad
\sin\theta = \frac{2\sqrt{tu - m_Z^4}}{\sqrt{(t + u)^2 - 4m_Z^4}}
\;,
\end{equation}
\begin{align}\label{t_u_ZZ}
t &= - \frac{1}{2} \,\left[ (s - 2m_Z^2) - \sqrt{s(s -
4m_Z^2)}\cos\theta \right] ,
\nonumber \\
u &= - \frac{1}{2} \,\left[ (s - 2m_Z^2) + \sqrt{s(s -
4m_Z^2)}\cos\theta \right] .
\end{align}
We have derived the following helicity amplitudes of the $ZZ
\rightarrow \gamma\gamma$ process:
\begin{align}\label{M++++ZZ}
M_{++++}^{ZZ} &=
-\frac{4s_w^2}{c_w^2}\frac{1}{f_a^2}\,\frac{s^{3/2}\sqrt{s -
4m_Z^2}}{s - m_a^2 + im_a \Gamma_a} +
\frac{s_w^2}{c_w^2}\frac{1}{f_a^2} \,s\!\left( \sqrt{s} - \sqrt{s -
4m_Z^2} \right)^{\!\!2}
\nonumber \\
&\times \left[ \frac{(1 + \cos\theta)^2}{t - m_a^2 + im_a \Gamma_a}
+ \frac{(1 - \cos\theta)^2}{u - m_a^2 + im_a \Gamma_a} \right] ,
\end{align}

\begin{align}\label{M+++-ZZ}
M_{+++-}^{ZZ} &= -\frac{4s_w^2}{c_w^2}\frac{1}{f_a^2}\,m_Z^2 s
(\sin\theta)^2
\nonumber \\
&\times \left[ \frac{1}{t - m_a^2 + im_a \Gamma_a} + \frac{1}{u -
m_a^2 + im_a \Gamma_a} \right] ,
\end{align}

\begin{align}\label{M+-++ZZ}
M_{+-++}^{ZZ} &= \frac{4s_w^2}{c_w^2}\frac{1}{f_a^2}\,m_Z^2 s
(\sin\theta)^2
\nonumber \\
&\times \left[ \frac{1}{t - m_a^2 + im_a \Gamma_a} + \frac{1}{u -
m_a^2 + im_a \Gamma_a} \right] ,
\end{align}

\begin{align}\label{M+-+-ZZ}
M_{+-+-}^{ZZ} &= -\frac{s_w^2}{c_w^2}\frac{1}{f_a^2}\,(1 +
\cos\theta)^2
\nonumber \\
&\times \left\{ \frac{\left[ s - \sqrt{s(s - 4m_Z^2)} \right]^2}{t -
m_a^2 + im_a \Gamma_a} + \frac{\left[ s + \sqrt{s(s - 4m_Z^2)}
\right]^2}{u - m_a^2 + im_a \Gamma_a} \right\} ,
\end{align}

\begin{align}\label{M++--ZZ}
M_{++--}^{ZZ} &=
\frac{4s_w^2}{c_w^2}\frac{1}{f_a^2}\,\frac{s^{3/2}\sqrt{s -
4m_Z^2}}{s - m_a^2 + im_a \Gamma_a} +
\frac{s_w^2}{c_w^2}\frac{1}{f_a^2}\,\left[ s + \sqrt{s(s - 4m_Z^2)}
\right]^2
\nonumber \\
&\times \left[ \frac{(1 - \cos\theta)^2}{t - m_a^2 + im_a \Gamma_a}
+ \frac{(1 + \cos\theta)^2}{u - m_a^2 + im_a \Gamma_a} \right] ,
\end{align}

\begin{align}\label{M+--+ZZ}
M_{+--+}^{ZZ} &= -\frac{s_w^2}{c_w^2}\frac{1}{f_a^2}(1 -
\cos\theta)^2
\nonumber \\
&\times \left\{ \frac{\left[ s + \sqrt{s(s - 4m_Z^2)} \right]^2}{t -
m_a^2 + im_a \Gamma_a} + \frac{\left[ s - \sqrt{s(s - 4m_Z^2)}
\right]^2}{u - m_a^2 + im_a \Gamma_a} \right\} ,
\end{align}

\begin{align}\label{M0+++ZZ}
M_{0+++}^{ZZ} &=
\frac{4i}{\sqrt{2}}\frac{s_w^2}{c_w^2}\frac{1}{f_a^2} \,m_Z s \left(
\sqrt{s} - \sqrt{s - 4m_Z^2} \right) \sin\theta
\nonumber \\
&\times \left [\frac{(1 + \cos\theta)}{t - m_a^2 + im_a \Gamma_a} -
\frac{(1 - \cos\theta)}{u - m_a^2 + im_a \Gamma_a} \right] ,
\end{align}

\begin{align}\label{M0++-ZZ}
M_{0++-}^{ZZ} &=
-\frac{4i}{\sqrt{2}}\,\frac{s_w^2}{c_w^2}\frac{1}{f_a^2} \,m_Z
\sqrt{s} (1 - \cos\theta)\sin\theta
\nonumber \\
&\times \left[ \frac{ s + \sqrt{s(s - 4m_Z^2)}}{t - m_a^2 + im_a
\Gamma_a} + \frac{ s - \sqrt{s(s - 4m_Z^2)}}{u - m_a^2 + im_a
\Gamma_a} \right] ,
\end{align}

\begin{align}\label{M0-++ZZ}
M_{0-++}^{ZZ} &=
\frac{4i}{\sqrt{2}}\,\frac{s_w^2}{c_w^2}\frac{1}{f_a^2} \,m_Z
\sqrt{s} \left[ s + \sqrt{s(s - 4m_Z^2)} \right] \sin\theta
\nonumber \\
&\times \left[ \frac{1 - \cos\theta}{t - m_a^2 + im_a \Gamma_a} -
\frac{1 + \cos\theta}{u - m_a^2 + im_a \Gamma_a} \right] ,
\end{align}

\begin{align}\label{M0+-+ZZ}
M_{0+-+}^{ZZ} &=
\frac{4i}{\sqrt{2}}\,\frac{s_w^2}{c_w^2}\frac{1}{f_a^2} \,m_Z
\sqrt{s} (1 + \cos\theta)\sin\theta
\nonumber \\
&\times \left[ \frac{ s - \sqrt{s(s - 4m_Z^2)}}{t - m_a^2 + im_a
\Gamma_a} + \frac{ s + \sqrt{s(s - 4m_Z^2)}}{u - m_a^2 + im_a
\Gamma_a} \right] ,
\end{align}

\begin{align}\label{M0+--ZZ}
M_{0+--}^{ZZ} &=
-\frac{4i}{\sqrt{2}}\,\frac{s_w^2}{c_w^2}\frac{1}{f_a^2} \,m_Z
\sqrt{s} \left[ s + \sqrt{s(s - 4m_Z^2)} \right] \sin\theta
\nonumber \\
&\times \left[ \frac{1 - \cos\theta}{t - m_a^2 + im_a \Gamma_a} -
\frac{1 + \cos\theta}{u - m_a^2 + im_a \Gamma_a} \right] ,
\end{align}

\begin{align}\label{M00++ZZ}
M_{00++}^{ZZ} &= -\frac{8s_w^2}{c_w^2}\frac{1}{f_a^2} \,m_Z^2 s
(\sin\theta)^2
\nonumber \\
&\times \left[ \frac{1}{t - m_a^2 + im_a \Gamma_a} + \frac{1}{u -
m_a^2 + im_a \Gamma_a} \right] ,
\end{align}

\begin{align}\label{M00+-ZZ}
M_{00+-}^{ZZ} &= -\frac{8s_w^2}{c_w^2}\frac{1}{f_a^2} \,m_Z^2 s
(\sin\theta)^2
\nonumber \\
&\times \left[ \frac{1}{t - m_a^2 + im_a \Gamma_a} + \frac{1}{u -
m_a^2 + im_a \Gamma_a} \right] .
\end{align}
Other helicity amplitudes
$M_{\lambda_1\lambda_2\lambda_3\lambda_4}^{ZZ}$ can be obtained
using relation \cite{Gounaris:2000}
\begin{equation}\label{parity_relation_ZZ}
M_{\lambda_1\lambda_2\lambda_3\lambda_4}^{ZZ} = (-1)^{\lambda_1 -
\lambda_2} M_{-\lambda_1-\lambda_2-\lambda_3-\lambda_4}^{ZZ} \;,
\end{equation}
where $\lambda_1, \lambda_2$ are helicities of the colliding $Z$
bosons.




\end{document}